\documentclass[%
 amsmath,amssymb,
 aps,
]{revtex4-2}

\usepackage{graphicx}
\usepackage{dcolumn}
\usepackage{bm}
\usepackage[normalem]{ulem}
\usepackage[dvipsnames]{xcolor}

\begin{document}


\title{Modelling Surface Segregation in Compositionally Complex Alloys with Ab-Initio Accuracy}

\author{Alberto Ferrari}
\email{a.ferrari-1@tudelft.nl}
\affiliation{Materials Science and Engineering,\\Delft University of Technology,\\2628CD Delft, The Netherlands}

\author{Vadim Sotskov}%
\affiliation{Skolkovo Institute of Science and Technology,\\Skolkovo Innovation Center,\\ Bolshoy Bulvar 30c1, 143026 Moscow, Russia
}%

\author{Alexander V. Shapeev}%
\affiliation{Skolkovo Institute of Science and Technology,\\Skolkovo Innovation Center,\\ Bolshoy Bulvar 30c1, 143026 Moscow, Russia
}%

\author{Fritz K\"ormann}
\email{f.koermann@mpie.de}
\affiliation{Materials Science and Engineering,\\Delft University of Technology,\\2628CD Delft, The Netherlands}

\affiliation{Max-Planck-Institut f\"ur Eisenforschung GmbH,\\40237 D\"usseldorf, Germany}

\date{\today}

\begin{abstract}
Compositionally complex alloys or concentrated solid solutions are the latest frontier in catalyst design, but mixing different elements in one catalyst may result in surface segregation. Atomistic simulations can predict segregation patterns, but standard approaches based on mean-field models, cluster expansion, or classical interatomic potentials are often limited for the description of multicomponent alloys. We  present machine learning potentials that can describe surface segregation with near DFT accuracy. The method is used to study a complex Co-Cu-Fe-Mo-Ni quinary alloy. For this alloy, an unexpected segregation of Co, which has a relatively high surface energy, is observed. We rationalize this surprising mechanism in terms of simple transition-metal chemistry. 
\end{abstract}

\maketitle

\section{Introduction}
\begin{figure*}
\begin{center}
\includegraphics[width=0.8\textwidth]{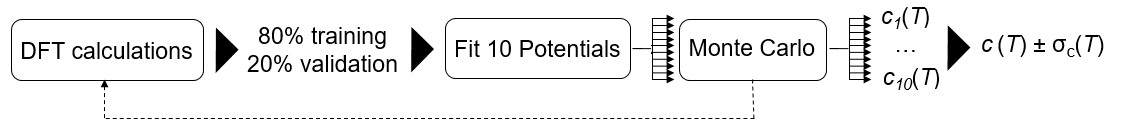}
\end{center}
\caption{The workflow employed in this work: the energies of a few hundred structures are computed with DFT; this database is divided into training and validation sets; ten LRPs are fitted; Monte Carlo simulations are carried out for each potential; the final composition of the surface is obtained by averaging the results of the ten simulations. Snapshots from the Monte Carlo simulations may be added to the DFT database for reinforcement.}
\label{workflow}
\end{figure*}

\begin{figure}[b]
\begin{center}
\includegraphics[width=0.3\textwidth]{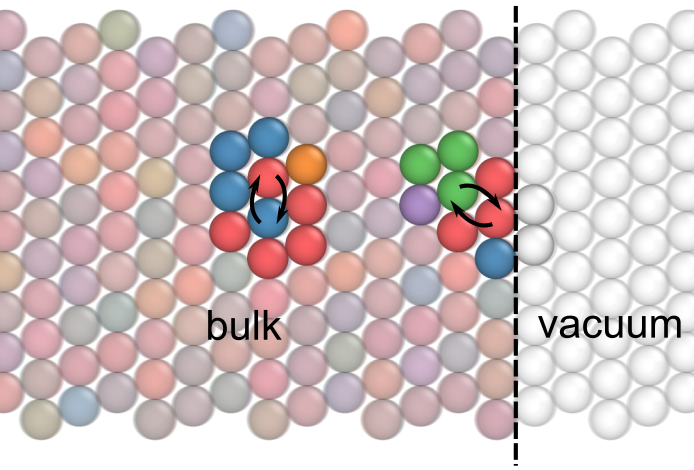}
\end{center}
\caption{Schematics of the fcc(111) slab used in the Monte Carlo simulations. The vacuum is treated as an additional species, here in white, so that the supercell maintains a fcc structure. Two possible Monte Carlo moves are pictured by arrows. Opaque atoms indicate the swapping atoms and their nearest neighbors.}
\label{structure}
\end{figure}

Compositionally complex alloys (CCAs), with three or more metals in high concentration, could be  game changers in heterogeneous catalysis, showing much higher catalytic activities than standard rule-of-mixtures would predict \cite{yao2018carbothermal,loffler2018discovery,loffler2019toward,batchelor2019high,sun2021high}. Compositional complexity, however, makes it hard to understand the catalytic processes of these alloys in terms of atomistic mechanisms. In fact, the kinetics of a chemical reaction strongly depend on the composition of a catalyst's surface, which, for multicomponent alloys, may be very different from the bulk: one or more elements may have a low surface energy and/or a weak binding to the other elements (high chemical potential) and hence segregate to the surface \cite{wynblatt2019modeling,ferrari2020surface,ferrari2020frontiers,ferrari2021design,ruban2021segregation,chatain2021surface}. As a result, the catalytic activity of an alloy is ultimately determined by surface segregation, rather than directly by the nominal composition.

Surface segregation can be understood and controlled with atomistic simulations. If segregation enthalpies $\Delta H_\text{seg}$ are known, the surface composition may be predicted with mean-field models, such as the Langmuir-McLean equation \cite{mclean1957grain}.
Such mean-field models could be extended to multicomponent alloys \cite{ruban2021segregation}, but may fail when the bonding between certain pairs of elements is much  stronger than others and a significant degree of short-range ordering is present. 

Monte Carlo simulations could directly incorporate eventual ordering effects. These simulations though require an accurate description of the interatomic interactions, which is difficult to obtain for multicomponent alloys. The most popular methods to parameterize these interactions for binary and ternary alloys are cluster expansion \cite{sanchez1984generalized,drautz2001spontaneous} and classical potentials \cite{daw1984embedded,baskes1992modified,foiles1987calculation}, both of which are, however, often impractical or not sufficiently accurate for alloys with more than three components because of the large number of parameters to be fitted. Effective interactions in slabs have also been computed with a perturbative approach based on the Coherent Potential Approximation \cite{soven1967coherent, gyorffy1972coherent}, called Generalized Perturbation Method (GPM) \cite{ducastelle1976generalized, schonfeld2019local,ruban2007theoretical}, but these calculations do not distinguish different local chemical environments and cannot incorporate relaxation energies.

In contrast, machine learning interatomic potentials have recently been shown to accurately describe the complex interactions occurring in bulk CCAs \cite{kostiuchenko2019impact,grabowski2019ab,jafary2019applying,kostiuchenko2020short,kormann2021b2,gubaev2021finite}. Among other formalisms \cite{hart2021-review-ML-for-alloys}, low-rank potentials (LRPs) \cite{shapeev2017accurate,kostiuchenko2019impact}, a type of \textit{on lattice} interaction model, have shone in terms of accuracy and efficiency in the description of local ordering effects in CCAs. LRPs essentially approximate the $m \times m \times ... \times m$, $n$-dimensional tensor that describes the interaction among $n$ atoms of species $1, ..., m$ with a tensor with a low rank $\bar{r}$, thereby reducing the number of free parameters from $m^n$ to $\text{O}\big(mn\bar{r}^2\big)$. LRPs are defined by specifying only two hyperparameters: the maximum rank $\bar{r}$ and the number of interacting neighbors $n$.

In this work we show that LRPs can accurately describe surface segregation mechanisms in CCAs. After benchmarking our approach against  literature data for the well-studied AgPd alloy, we compute the surface composition of the  $\text{Co}_{25}\text{Cu}_{10}\text{Fe}_{10}\text{Mo}_{45}\text{Ni}_{10}$ alloy. This alloy was experimentally shown to catalyze very efficiently the $\text{NH}_3$ decomposition reaction \cite{xie2019highly}, which may become key for the emerging hydrogen economy. Previous Monte Carlo calculations \cite{xie2019highly} on $\text{Co}_{25}\text{Cu}_{10}\text{Fe}_{10}\text{Mo}_{45}\text{Ni}_{10}$ nanoparticles with a Modified Embedded Atom Method (MEAM) potential \cite{baskes1992modified} showed no significant surface segregation. This appears surprising given the heterogeneity of the constituents in the alloy and may be due to the short annealing time employed in those calculations to simulate kinetic hindering. Based on the accurate machine learning interatomic potential, our simulations point out a significant Co segregation instead. This finding is corroborated by a simple $d$-band model. 

We focused on the fcc(111) surfaces of AgPd and $\text{Co}_{25}\text{Cu}_{10}\text{Fe}_{10}\text{Mo}_{45}\text{Ni}_{10}$ because these are the most closely packed. To determine the temperature dependence of the surface concentration $c(T)$ of an element,  we employed the workflow in Fig.~\ref{workflow}. We first calculated the total energy of a few hundred slabs with Density Functional Theory (DFT) including magnetism and atomic relaxations. We independently fitted ten LRPs on the training set (80\% of the slabs) by minimizing the mean-square error on the energy and then validated our fits on the remaining 20\% of the slabs. The training and validation sets were shuffled for each potential. With the parametrized LRPs, we performed Monte Carlo simulations to determine the composition of the surface. We compared the outcomes for the ten potentials and took the average and standard deviation of the results. If the standard deviation was too large or if the potentials gave inconsistent results, snapshots were extracted from the Monte Carlo simulations and calculated with DFT to reinforce the fitting. For the LRPs, we fixed the interaction range to include only the nearest neighbors ($n=13$) and tested different maximum ranks $\bar{r}$. Independent calculations for the bulk $\text{Co}_{25}\text{Cu}_{10}\text{Fe}_{10}\text{Mo}_{45}\text{Ni}_{10}$ were performed using the same methodology.
\textcolor{black}{For each alloy a different potential has been fitted. We note that the methodology itself does not have a limitation with regards to the compositional range – a machine learning potential can be trained on any compositional range.  However, a particular instance of a trained machine learning potential is typically limited to the compositions it was trained on and the extrapolation error is \textit{a priori} not known.}

For the Monte Carlo simulations we adopted the setup in Fig.~\ref{structure}. To maintain the symmetry of an fcc lattice, we treated vacuum as an additional species. The PdAg and $\text{Co}_{25}\text{Cu}_{10}\text{Fe}_{10}\text{Mo}_{45}\text{Ni}_{10}$ then effectively became a ternary ($m=3$) and a senary ($m=6$) alloy, respectively. We considered nearest-neighbor swaps as our Monte Carlo moves, but we excluded swaps involving the ``vacuum atoms". For bulk calculations discussed later, a similar setup was chosen. For the technical details of the DFT and Monte Carlo simulations we refer to Sec.~\ref{methods}.  Our simulations fully account for relaxations effects at the surface and in the bulk.
We did not include atomic vibrations and volume changes in our Monte Carlo simulations because the differences of the vibrational and volume contributions to the free energy are usually negligible compared to segregation energy differences \cite{ferrari2020surface,ferrari2021design}.

\section{Results}
\subsection{Benchmark for AgPd}
\begin{figure}[h]
\begin{center}
\includegraphics[width=0.5\textwidth]{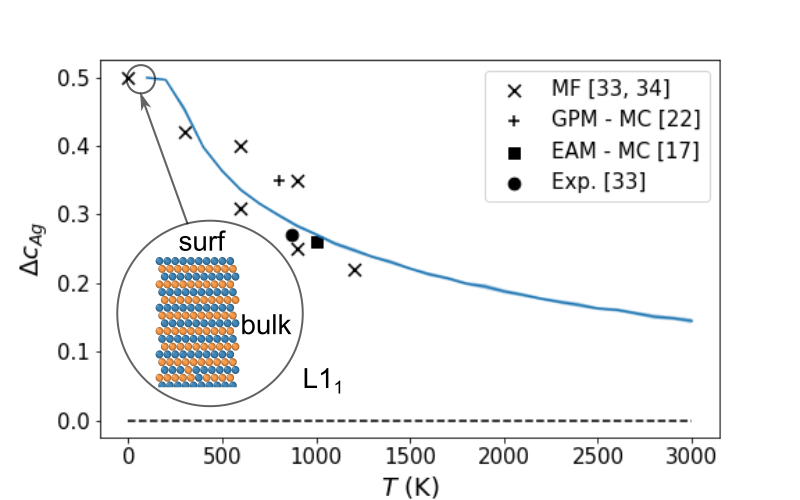}
\end{center}
\caption{Excess concentration of Ag on the surface layer of AgPd as a function of temperature. Points are from literature: MF = Mean Field model with segregation energies calculated with DFT; GPM - MC = Monte Carlo with interaction energies from the Generalized Perurbation Method; EAM - MC = Monte Carlo with an Embedded Atom Method potential; Exp. = experimental (Auger Electron Spectroscopy). The inset shows a snapshot from a Monte Carlo simulation at 100 K, where the precipitation of the $\text{L}1_1$ phase is evident.}
\label{PdAg}
\end{figure}

\begin{table}[h]
    \centering
    \begin{tabular}{cc}
    \hline
         & Surface Energy (mJ/m$^2$) \\
    \hline
        Ag & 760 \\
        Pd & 1360 \\
    \hline
        Cu & 1340  \\
        Ni & 1920  \\
        Co & 2110  \\
        Fe & 2450  \\
        Mo & 2780  \\
    \hline
    \end{tabular}
    \caption{Formation energies of surfaces for the pure elements from the Materials Project database \cite{Jain2013}. The surface with the lowest formation energy for the most stable structure was considered for each element. }
    \label{surface_energies}
\end{table}

For the AgPd system we reached training and validation errors as low as 0.4 meV/at.~for a set of LRPs with maximum rank $\bar{r}=3$. The resulting surface composition can be deduced from Fig.~\ref{PdAg}, which shows the excess concentration of Ag on the surface layer with respect to the bulk averaged over our ensemble of ten LRPs. Surface segregation is very prominent in this case, since there is Ag enrichment even at a temperature as high as 3000 K, way beyond the melting temperature (if it were possible to keep the alloy from melting; note that the utilized on-lattice approach obviously cannot capture the solid-liquid transition). The segregation is mainly driven by the much lower surface energy of Ag with respect to Pd (see Tab.~\ref{surface_energies}).

The obtained results compare very well to previous investigations, where the Ag concentration was estimated with a mean-field model
\cite{ropo2005segregation,ropo2006chemical} as well as with Monte Carlo simulations based on interactions determined with the Generalized Perturbation Method \cite{ruban2007theoretical} or with an Embedded Atom Method potential \cite{foiles1987calculation}. Good agreement is also found with the experimental data of Ref.~\onlinecite{reniers1994surface} refined by Ref.~\onlinecite{ropo2005segregation}.

In agreement with other works \cite{muller2001first,sluiter2006ab,ruban2007theoretical} we observed the emergence of long-range ordering at low temperature, with Ag and Pd layers alternating along the $[111]$ direction (see inset of Fig.~\ref{PdAg}). This is due to the known formation of the $\text{L}1_1$ phase in this alloy. Note that this phase was not part of the training set of our LRPs, but it was nevertheless correctly identified as ground state. 

\subsection{The $\text{Co}_{25}\text{Cu}_{10}\text{Fe}_{10}\text{Mo}_{45}\text{Ni}_{10}$ alloy}
\begin{table}[h]
    \begin{center}
    \begin{tabular}{c|cc|cc}
    \hline
         & \multicolumn{2}{c|}{Train Err. (meV/at.)} & \multicolumn{2}{c}{Valid.~Err. (meV/at.)} \\
         $\bar{r}$& Surf. & Bulk & Surf. & Bulk \\
         \hline
        3 & 4.2 & 3.5 & 4.5 & 4.9\\
        4 & 3.5 & 2.7 & 3.9 & 4.3\\
        5 & 3.0 & 2.1 & 4.1 & 4.2\\
    \hline
    \end{tabular}
    \end{center}
    \caption{Average fitting and validation root-mean-squared errors for the LRPs for the $\text{Co}_{25}\text{Cu}_{10}\text{Fe}_{10}\text{Mo}_{45}\text{Ni}_{10}$ system for the surface segregation and bulk simulations.}
    \label{error_LRP}
\end{table}

\begin{figure}
\begin{center}
\includegraphics[width=0.5\textwidth]{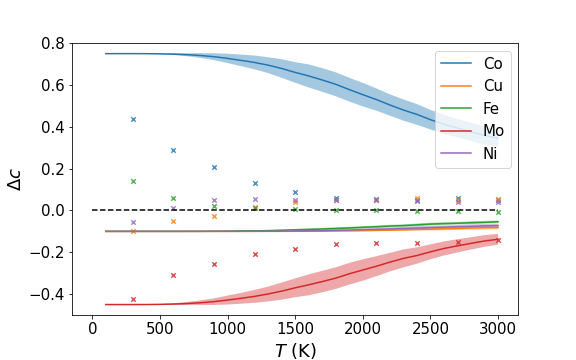}
\end{center}
\caption{Excess concentration of the component elements on the surface layer of $\text{Co}_{25}\text{Cu}_{10}\text{Fe}_{10}\text{Mo}_{45}\text{Ni}_{10}$ as a function of temperature. The shaded areas indicate an interval of one standard deviation obtained over ten fitted potentials. Dashed lines correspond to the nominal bulk concentrations. Points are from a simple $d$-band model (see Sec.~\ref{dmodel}).}
\label{CoCuFeMoNi}
\end{figure}

For the $\text{Co}_{25}\text{Cu}_{10}\text{Fe}_{10}\text{Mo}_{45}\text{Ni}_{10}$ system we tested different values of the maximum rank $\bar{r}$, as reported in Tab.~\ref{error_LRP}. We found that optimal values for this alloy are $\bar{r}=4$ or $\bar{r}=5$, yielding training and validation errors around 3--4 meV/at.; in the following we report the results obtained with $\bar{r}=5$.

The excess concentrations of the component elements on the surface of $\text{Co}_{25}\text{Cu}_{10}\text{Fe}_{10}\text{Mo}_{45}\text{Ni}_{10}$ with respect to the bulk are displayed as solid lines in Fig.~\ref{CoCuFeMoNi} as a function of temperature. The shaded areas indicate an interval of one standard deviation obtained over ten fitted potentials. In contrast to Ref.~\onlinecite{xie2019highly}, we see a large Co segregation in this system that, similarly as for the AgPd system, persists up to very high temperature.

\section{Discussion}
\subsection{Why does Co segregate?}
\begin{figure*}
\begin{center}
\includegraphics[width=1.0\textwidth]{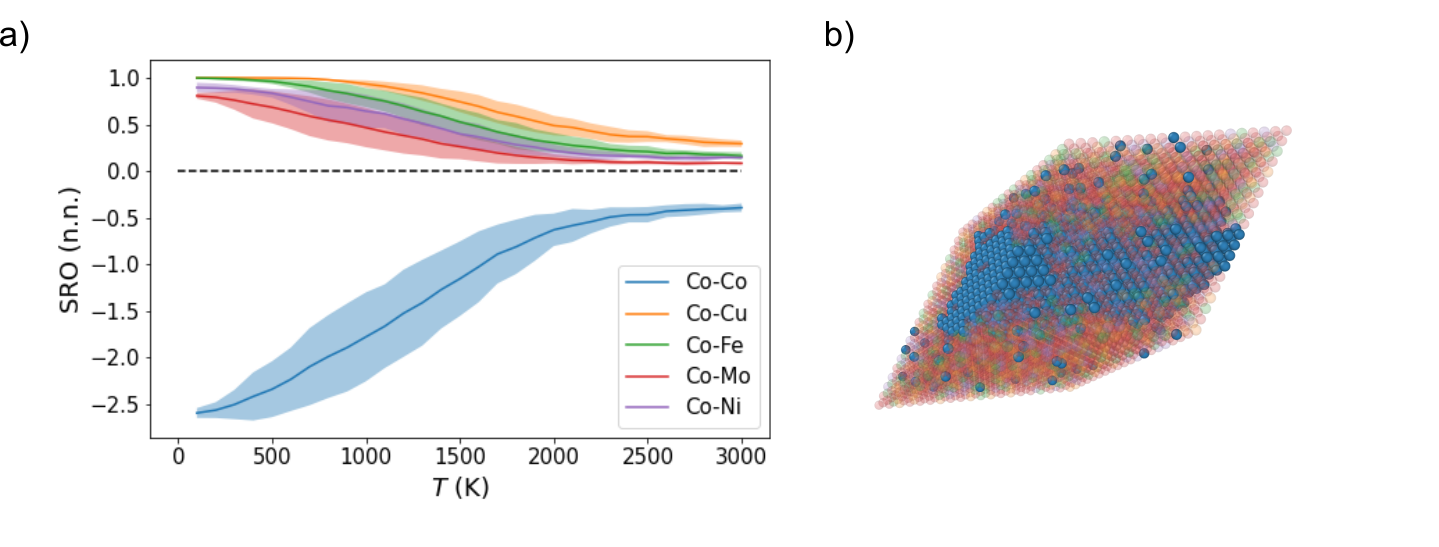}
\end{center}
\caption{a) Bulk short-range order parameters for the first neighbor shells for the element pairs involving Co as a function of temperature. A negative value indicates favoured pairs. The shaded areas denote an interval of one standard deviation obtained over ten fitted potentials. b) A snapshot of the bulk $\text{Co}_{25}\text{Cu}_{10}\text{Fe}_{10}\text{Mo}_{45}\text{Ni}_{10}$ at 1000 K. Co atoms are highlighted in blue.}
\label{SRO}
\end{figure*}


The segregation of Co is at first quite unexpected, since it is not the element with the lowest surface energy among the five components (see Tab.~\ref{surface_energies}). In fact, this phenomenon originates in the bulk: we observed that in this alloy Co has a low affinity for the other elements and at low temperature it even shows a marked tendency for demixing. The surface acts therefore as a sink for the Co that escapes from the bulk.

To analyze in detail this aspect, we fitted another set of LRPs on bulk fcc structures for $\text{Co}_{25}\text{Cu}_{10}\text{Fe}_{10}\text{Mo}_{45}\text{Ni}_{10}$, with a procedure analogous to that used for the (111) surface. The average training and validation errors for this new set of LRPs are reported in Tab.~\ref{error_LRP}. With the fitted potentials, we studied the bonding and eventual ordering modes in the bulk. The tendency of Co demixing in the bulk is confirmed by the analysis of the short-range order (SRO) parameter between nearest neighbor pairs, computed as
\begin{equation}
    \text{SRO}_{ij}=1-\frac{p_{ij}}{c_ic_j},
\end{equation}
where $p_{ij}$ is the observed probability of finding atom $j$ in the nearest neighborhood of $i$, and $c_i$ and $c_j$ are the bulk concentrations.
As shown in Fig.~\ref{SRO}a), the SRO parameter for the Co-Co pair is strongly negative, indicating overall attraction, whereas that for the other pairs is positive, indicating repulsion. According to our ensembles of potentials, the preferential bonding of Co to itself triggers even the precipitation of Co-rich domains at low temperature, as seen in Fig.~\ref{SRO}b). In this system, the segregation is therefore not driven by the lowest surface energy, but by the highest chemical potential (weakest bonding to the other elements).

\subsection{Co segregation emerges from a $d$-band effect}
\label{dmodel}
The mechanism that causes the segregation of Co descends from a purely electronic effect and can be understood in terms of  the so-called Friedel model \cite{sutton1993electronic,ruban1999surface}, that assumes a rectangular density of $d$-states with bandwidth $w$. According to this model, the binding energy of an alloy with $d$-band filling $f$ is
\begin{equation}
    E_\text{bind}=-5wf(1-f).
    \label{friedel}
\end{equation}

To demonstrate that for the $\text{Co}_{25}\text{Cu}_{10}\text{Fe}_{10}\text{Mo}_{45}\text{Ni}_{10}$ alloy the segregation of Co emerges from this simple $d$-band effect, we performed the same Monte Carlo calculations as before but with a simple pairwise interatomic potential that  depends only on the $d$-band fillings and $d$-bandwidths of the elements. We consider nearest-neighbor interactions only of type 
\begin{equation}
    V_{ij}=-\frac{5}{6}w f_{ij} (1-f_{ij}),
    \label{d-band-model1}
\end{equation}
where
\begin{equation}
    f_{ij}=\frac{1-\sqrt{|1-2(f_i+f_j-2f_if_j)|}}{2},
    \label{d-band-model2}
\end{equation}
with $f_i$ and $f_j$ the band fillings of the pure elements and $w$ an average bandwidth (see Sec.~\ref{methods} for the details). Eqs.~\ref{d-band-model1} and \ref{d-band-model2} were derived so that the binding energy of a binary alloy reduces to Eq.~\eqref{friedel}.

The excess concentrations of the component elements on the surface of $\text{Co}_{25}\text{Cu}_{10}\text{Fe}_{10}\text{Mo}_{45}\text{Ni}_{10}$ obtained from this simple $d$-band model are shown with points in Fig.~\ref{CoCuFeMoNi}. The segregation trend observed with this model is qualitatively similar to that obtained with the LRP simulations, i.e.~there is a clear Co enrichment on the surface. For this simplified model we also notice a hint of a possible Fe segregation at low temperature. 

Additional calculations with the $d$-band model, detailed in the Supplementary Material, show that the Co demixing and segregation depend strongly on the nominal composition of the bulk and that additions of Ni or Cu cause the segregation of these elements instead of Co. These results explain the segregation trend obtained with the LRPs with a $d$-band filling effect and fully validate our predictions in terms of fundamental transition-metals chemistry. 

\bigskip

In summary, we investigated surface segregation mechanisms in a binary and a quinary alloy with machine-learning interatomic potentials. We observed a very good agreement with the literature for the binary alloy, but a previously unmentioned Co segregation in the quinary alloy. We motivated this result with a high bulk chemical potential of this element and justified it with a simple, parameter-free $d$-band model. The present work highlights the potential of  machine-learning interatomic potentials for simulating ordering and surface segregation  in CCAs.

\section{Methods}
\label{methods}

\subsection{DFT database}
To fit the LRPs for AgPd we employed a total of 200 slabs with size $4\times4\times7$, of which 100 are random configurations and 100 were Monte Carlo snapshots added later for reinforcement. For $\text{Co}_{25}\text{Cu}_{10}\text{Fe}_{10}\text{Mo}_{45}\text{Ni}_{10}$ we instead employed a total of 591 slabs,  of which 507 with size $2\times2\times10$ and 84 with size $4\times4\times10$ (we mixed different supercell sizes to avoid possible spurious periodic interactions). The $4\times4\times10$ slabs and 100 of the $2\times2\times10$ slabs were random, the rest were extracted from Monte Carlo and added later to the fitting database. We also included ordered configurations where each element completely segregates. To fit the LRPs for the bulk $\text{Co}_{25}\text{Cu}_{10}\text{Fe}_{10}\text{Mo}_{45}\text{Ni}_{10}$ we employed 349 structures with size $5\times4\times4$, of which 138 were random and 211 were extracted from Monte Carlo. For the structures and energies used to fit the potentials see \url{https://github.com/AlbertoFerrari8/Segregation_LRP}.

For the DFT calculations we employed the {\sc vasp} 5.4 package \cite{kresse1993ab,kresse1996efficiency,kresse1996efficient} with Projector Augmented Wave potentials \cite{blochl1994projector,kresse1999ultrasoft}. We treated the exchange-correlation functional with the PBE approximation \cite{perdew1996generalized}. The energy cutoff was set to 400 eV and the width of the Methfessel-Paxton function \cite{methfessel1989high} for the electronic smearing to 0.1 eV. The $k$-point meshes were centered around $\Gamma$ and the points were distributed according to the Monkhorst-Pack grids \cite{baldereschi1973mean,monkhorst1976special} with a linear density of 0.1 $2\pi/\text{\AA}$. The calculations for $\text{Co}_{25}\text{Cu}_{10}\text{Fe}_{10}\text{Mo}_{45}\text{Ni}_{10}$ were spin-polarized. The initial magnetic moments were set to 2.0, 0.0, 3.0, 0.0, and 1.0 $\mu_\text{B}$. Around 20 \AA~of vacuum prevented the interaction between periodic slabs. For all supercells, we relaxed the atomic positions until the interatomic forces were less than 0.05 eV/\AA. The lattice parameter was fixed to 4.033 \AA ~or AgPd (the equilibrium value for a random alloy at 0 K) and to 3.776 \AA~for $\text{Co}_{25}\text{Cu}_{10}\text{Fe}_{10}\text{Mo}_{45}\text{Ni}_{10}$ (as in Ref.~\cite{xie2019highly}).

\subsection{Monte Carlo simulations}

The surface composition was obtained as an average of the two surfaces of a $10\times10\times40$ slab, whereas for the bulk simulations we employed a $20\times20\times20$ supercell. The Monte Carlo simulations were performed in the canonical ensemble. A Monte Carlo move consisted in the swap of two nearest neighbor atoms. Starting from a random configuration, we first performed $1\cdot10^{2}$ steps/atom for equilibration at 5000 K and then began averaging, first at 5000 K and then decreasing the temperature by 100 K at each cycle. Averages were taken on 100 snapshots $1\cdot10^{2}$ steps/atom apart, so that $1\cdot10^{4}$ steps/atom were taken in total for each temperature. Only two vacuum layers were considered in the Monte Carlo simulations for efficiency.

\subsection{$d$-band model}
The $d$-band fillings for the pure elements were computed by integrating the $d$-density of states in the fcc structure up to the Fermi level. These calculations were carried out with the exact muffin-tin orbital method \cite{vitos2000application, vitos2001total, vitos2007computational}. The fillings take the following values: Co: 0.76, Mo: 0.47, Fe: 0.65, Ni: 0.87, Cu: 0.96. We took the value of 4.2 eV for the bandwidth $w$, estimated as a weighted average of the bandwidths of Co, Fe, Ni, and Cu in the fcc structure at the lattice parameter of 3.776 \AA.

\section*{Acknowledgements}

AF and FK were supported by the Nederlandse Organisatie voor Wetenschappelijk Onderzoek (NWO) [VIDI Grant No.~15707]. Part of the calculations were conducted on the Dutch national e-infrastructure with the support of SURF Cooperative.
VS and AVS were supported by the Russian Science Foundation (Grant No 18-13-00479, \url{https://rscf.ru/project/18-13-00479/}).

\section*{Data Availability}

Derived data supporting the findings of this study are available from the corresponding author upon request.

\section*{Author Contributions}
AF and FK designed the research and performed the DFT calculations. VS and AVS developed the code for the LRP parametrization and Monte Carlo simulations. AF performed the fitting, the Monte Carlo simulations, and the post-processing. All authors discussed the results and wrote the article.

\section*{Competing Interests}
The authors declare no competing interests.

\bibliography{main.bib}

\end{document}